\documentclass[letterpaper]{article}

\usepackage{amssymb}
\usepackage{amsmath}
\usepackage{amsthm}
\usepackage[labelsep=period,skip=6pt,labelfont=bf]{caption}
\usepackage[usenames,dvipsnames]{xcolor}
\usepackage{enumerate}
\usepackage{fixmath}
\usepackage[letterpaper,nohead,nomarginpar,centering,margin=3cm]{geometry}
\usepackage{graphicx}
\usepackage{hyperref}
\usepackage{natbib}
\usepackage{ragged2e}
\usepackage{rotating}
\usepackage{setspace}
\usepackage{tabularx}
\usepackage[tiny]{titlesec}
\usepackage{doi}
\usepackage{mathptmx}
\usepackage{authblk}
\usepackage[symbol]{footmisc}
\usepackage{lineno}
\usepackage{xcolor}

\hypersetup{colorlinks,breaklinks,linkcolor=blue,urlcolor=blue,anchorcolor=blue,citecolor=blue}
\title{The Antarctic circumpolar wave and its seasonality:\\  Intrinsic traveling modes and ENSO teleconnections\\
\vspace {0.3cm}
{\large Short title: ACW: Intrinsic traveling modes and ENSO teleconnections}}

\author[1\footnote{Corresponding author. Address: Center for Atmosphere Ocean Science, Courant Institute of Mathematical Sciences, New York University, 251 Mercer Street
New York, NY,10012-1185, USA. E-mail: xinyang@cims.nyu.edu}]{Xinyang Wang}
\author[1]{Dimitrios Giannakis}
\author[2]{Joanna Slawinska}

\affil[1]{Center for Atmosphere Ocean Science, Courant Institute of Mathematical Sciences, New York University, New York, New York, USA}
\affil[2]{Department of Physics, University of Wisconsin-Milwaukee, Milwaukee, Wisconsin, USA}

\date{}

\begin{document}
\maketitle
%% ------------------------------------------------------------------------ %%
%
%  AUTHORS AND AFFILIATIONS
%
%% ------------------------------------------------------------------------ %%

% up to eight key words
\newpage

%% ------------------------------------------------------------------------ %%
%
%  ABSTRACT
%
%% ------------------------------------------------------------------------ %%

\begin{abstract}
    Interannual variability in the Southern Ocean is investigated via nonlinear Laplacian spectral analysis (NLSA), an objective eigendecomposition technique for nonlinear dynamical systems that can simultaneously recover multiple timescales from data with high skill. Applied to modeled and observed sea surface temperature and sea ice concentration data, NLSA recovers the wavenumber-2 eastward propagating signal corresponding to the Antarctic circumpolar wave (ACW). During certain phases of its lifecycle, the spatial patterns of this mode display a structure that can explain the statistical origin of the Antarctic dipole pattern. Another group of modes have combination frequencies consistent with the modulation of the annual cycle by the ACW. Further examination of these newly identified modes reveals that they can have either eastward or westward propagation, combined with meridional pulsation reminiscent of sea ice reemergence patterns in the Arctic. Moreover, they exhibit smaller-scale spatial structures, and explain more Indian Ocean variance than the primary ACW modes. We attribute these modes to teleconnections between ACW and the tropical Indo-Pacific Ocean; in particular, fundamental ENSO modes and their associated combination modes with the annual cycle recovered by NLSA. Another mode extracted from the Antarctic variables displays an eastward propagating wavenumber-3 structure over the Southern Ocean, but exhibits no strong correlation to interannual Indo-Pacific variability.

\end{abstract}

%% ------------------------------------------------------------------------ %%
%
%  TEXT
%
%% ------------------------------------------------------------------------ %%

\section{Introduction}

The dominant interannual mode of variability in the Southern Ocean, called the Antarctic Circumpolar Wave (ACW), plays an important role in the global climate, but its spatiotemporal structure, propagation characteristics, and associated physical mechanisms remain partly understood. In particular, a long-standing topic under investigation has been the role of tropical-extratropical forcing versus local atmosphere-ocean coupling in ACW dynamics. Historically, the ACW was first identified by \citet{WP1996} as an eastward-propagating signal in observed Antarctic sea surface temperature (SST), sea level pressure (SLP), sea-ice extent, and wind anomalies, with a periodicity of around 4--5 years and a zonal wavenumber-2 structure. Subsequently, \citet{WP1998} proposed an ENSO-ACW forcing mechanism. On the other hand, model studies have found wavenumber-3 patterns and associated them with an ocean-atmosphere coupling, suggesting the significance of local dynamics underlying the generation and sustenance of the ACW \citep{C1998, BC1999}. 
 
A diverse range of data analysis techniques have been employed to investigate the existence of wavenumber-2 \citep{YM2001, CN2001,V2003} and wavenumber-3 \citep{BC2001,V2003, P2004, Cerrone2017a,Cerrone2017b, CF2018} structures in ACW signals, lending support to the hypotheses that these signals are closely related to ENSO teleconnections \citep{WP1998, YM2001, BC2001, V2003, Cerrone2017a,Cerrone2017b} and local atmosphere-ocean coupling \citep{V2003,Cerrone2017a,Cerrone2017b, CF2018}, respectively. In particular, a number of studies \citep{BC2001,V2003,Cerrone2017a,Cerrone2017b} point to the fact that the dominant ACW structures can be explained as a combination of ENSO teleconnection, mediated by the Pacific South American (PSA) pattern, and extratropical zonal wavenumber-3 modes. 

Yet, there exist significant discrepancies in the ACW propagation features among studies employing different data analysis techniques. \citet{WP1998} recovered the eastward-propagating ACW by extended EOF analysis of interannual SST, SLP, and precipitable water anomalies. \citet{BC2001} also identified eastward-propagating ACW modes, of both wavenumber-2 and 3, via complex EOF analysis of SST data. In contrast, \citet{YM2001} found a dominant quasi-stationary wave called the Antarctic Dipole by applying conventional EOF analysis to SST and sea ice extent data, while \citet{PR2004} found that most of the Antarctic interannual variability can be explained by a standing wavetrain using a Fourier decomposition. As another example, the wavenumber-3 signal was identified as a standing oscillation in SLP anomalies by \citet{V2003}, versus an eastward-propagating mode in SST and 850hPa geopotential height field by \citet{Cerrone2017a}, with both studies using the multi-taper method with singular value decomposition (MTM-SVD) \citep[][]{MannPark99}. On the other hand, \citet{P2004} found via a space-time spectral analysis that it exhibits a westward propagation in sea level anomalies.

Arguably, such discrepancies in the identified ACW propagation characteristics are at least partly caused by data analysis techniques relying on subjective filtering to isolate the temporal and spatial scales of interest. For instance, \citet{YM2001} and \citet{PR2004} utilize a $ \sim 1 $ yr low-pass filter which is shorter than the interannual filters used in some other studies \citep[e.g.,][]{WP1996}. More broadly, classical EOF analysis is known to mix signals from distinct physical processes, hampering the physical interpretability of the results \citep{VonStorchZwiers02}. On the other hand, spectral estimation techniques are capable of identifying narrowband signal components, but generally depend on a number of subjective choices, such as windowing and background removal. Classical EOF and spectral estimation approaches may both have limited skill in extracting low-variance, yet dynamically significant, components of signals.  

In response, the aim of this study is to explore the temporal and spatial patterns of the ACW through the use of objective eigendecomposition techniques for dynamical systems requiring no preprocessing of the input data. Our main methodological tool is the  nonlinear Laplacian spectral analysis (NLSA) \citep{GiannakisMajda12a, GiannakisMajda12b, GiannakisMajda13}; a framework that combines aspects of ergodic theory with kernel methods from machine learning to recover intrinsic temporal and spatial patterns associated with the point frequency spectrum of the dynamical system generating the data. Among a number of climate dynamics applications, NLSA has previously been employed in a diagnostic study of Indo-Pacific SST variability on seasonal to decadal timescales  \citep[][hereafter, SG]{SlawinskaGiannakis17,GiannakisSlawinska18}. The hierarchy of modes recovered in this study includes the annual cycle and its harmonics, ENSO, the tropospheric biennial oscillation (TBO) \citep{Meehl97}  and the interdecadal Pacific oscillation (IPO) \citep{PowerEtAl99}. In addition, the method identified a family of combination modes representing the interaction between ENSO and the seasonal cycle \citep[][]{StueckerEtAl13}, and a new decadal mode called west Pacific multidecadal mode (WPMM) that was found to exhibit significant correlations with ENSO activity.  

Here, we apply this framework to analyze model and observational SST and sea ice concentration (SIC) data. In addition to the properties of the dominant interannual modes, which characterize the wavenumber-2 and 3 ACW, we examine the role of modulating relationships with the seasonal cycle in the ACW propagation characteristics. We also investigate ENSO-ACW teleconnections by comparing modes recovered from Indo-Pacific and Antarctic data. 

\section{\label{secData}Data and methods}

\subsection{Dataset description}

The data studied in this paper consist of monthly averaged SIC and SST data from a 1300-year control integration of the Community Climate System Model version 4 (CCSM4) \citep{GentEtAl11} and the Hadley Centre Sea Ice and Sea Surface Temperature data set (HadISST) \citep{RaynerEtAl03} from 1979 to 2017. The CCSM4 datasets utilize the model's native ocean grid of a $1^{\circ}$ nominal resolution. The HadISST data are on a uniform 1$6^{\circ}$ longitude-latitude grid. 

Compared to its predecessor (CSSM3), CCSM4 produces ENSO variability with a significantly more realistic frequency spectrum in the 3 to 6 yr$^{-1}$ frequency band, and its simulated annual cycle of Pacific SST anomalies is closer to observations \citep{GentEtAl11,DeserEtAl12b}. This improvement has been attributed to changes in the convective parameterization scheme, resulting in improved representation of organized convection---a significant source of bias in climate models \citep{SlawinskaEtAl15}. While CCSM4 has a positive ENSO amplitude bias \citep{DeserEtAl12b},  its Southern Ocean surface climatology, which is influenced by atmospheric circulation patterns such as the PSA (the ENSO response  in the southern mid to high latitudes)
%and southern annualar mode(SAM)
is in reasonably good agreement with reanalysis and observations \citep{Weijer2012}. The latter is important in the context of the present study, as we seek to characterize the relationship between ENSO and the ACW. In terms of sea ice, CCSM4 uses a delta-Eddington radiative transfer scheme \citep{BL2017} to incorporate snow, sea ice, melt ponds, and other absorbers, resulting in more realistic surface ice albedos and shortwave radiative transfer in the ice and overlying snowpack \citep{GentEtAl11}. The model also exhibits a more realistic wind stress driving the Antarctic Circumpolar Current (ACC), but nevertheless ACC transport in CCSM4 is higher than in observations (though to a lesser degree than CCSM3) \citep{Weijer2012}. Despite this bias, the model simulates adequately the main geometrical configuration of the ACC jets.  

In what follows, we perform analyses of SST and SIC anomalies in the latitude belt $45^{\circ}$S--$90^{\circ}$S to investigate the ACW's characteristics, and we also analyze SST anomalies in the Indo-Pacific longitude-latitude box $28^{\circ}$E--$70^{\circ}$W, $30^{\circ}$S--$20^{\circ}$N to study ACW-ENSO teleconnections. Note that these two domains are non-overlapping. Unlike previous studies, the data in this work are not subjected to any preprocessing procedure such as detrending, bandpass filtering, and removal of the seasonal cycle. This allows us to detect  interactions between dominant low-frequency (interannual) modes of variability and the annual cycle.  It will be shown below that these interactions capture distinct propagation characteristics, and are consistent between model and observational data. 

In addition to surface-level data, we study tropospheric processes associated with the ACW through reconstructions of 850 hPa geopotential height data from CCSM4, based on modes derived from the SST and SIC datasets described above. Such tropospheric processes include Rossby wave trains associated with the PSA pattern \citep{Hoskins1981,MH1998}, which originate from tropical deep convection \citep{SlawinskaEtAl14} and can mediate the ENSO--ACW teleconnection. Conversely, negative SIC and positive SST anomalies associated with the ACW drive poleward surface winds and deep convection associated with anomalous low-level diabatic cooling and mid- to upper-level diabatic heating in the troposphere \citep{W2002, W2004, W2006}, leading to teleconnections from the ACW to the subtropics. 

\subsection{Overview of the nonlinear Laplacian spectral analysis (NLSA) framework}

As in extended EOF analysis and singular spectrum analysis (SSA) \citep{GhilEtAl02}, NLSA operates on data in delay-embedding space. That is, the first step in NLSA is to transform a sequence of snapshots $ x_i \in \mathbb{R}^d $ of a climatic variable sampled at times $t_i $ (here, every month) at $ d $ spatial gridpoints, to a sequence $ X_0, X_1, \ldots, X_{N-1} $ of concatenated snapshots $X_i=(x_i, x_{i-1}, \dots, x_{i-q+1}) \in \mathbb{R}^{qd}$. Then, the delay-embedded data are used to compute the values $ K_{ij} = K(X_i, X_j) $ of a nonlinear kernel function, measuring the similarity between data points. The resulting $N \times N $ kernel matrix $ K = [ K ]_{ij }  $ is subsequently normalized to form a Markov matrix $ P $ using the procedure introduced in the diffusion maps algorithm \citep{CoifmanLafon06} for machine learning, and temporal patterns $ \phi_k \in \mathbb{R}^N $ are recovered by the eigenvectors of $P$. 

The NLSA temporal patterns can be thought of as nonlinear analogs of the principal components in EOF analysis and SSA. Once they have been computed, they can be used for spatiotemporal reconstruction as in those techniques \citep[e.g.,][]{GhilEtAl02,GiannakisMajda12b,SlawinskaGiannakis17}. More specifically, given an eigenvector $\phi_k$, and a target variable $y_i$ observed at the same time instances as $ x_i $, we first compute the spatiotemporal pattern  ${Y}^{(k)}= Y\phi_k\phi^T_k$ in lagged-embedding space,  where $ Y = ( Y_0, \ldots, Y_{N-1} ) $, and then project that pattern to a spatiotemporal pattern ${y}^{(k)} = ( y_0^{(k)}, \ldots, y_{N-1}^{(k)})$ in the physical space by averaging the diagonal entries along the blocks of ${Y}^{(k)}$ corresponding to the same timestamp in physical space. The sum of the $y^{(k)}$ from all eigenfunctions identically recovers $ y = ( y_0, \ldots, y_{N-1} ) $.  Further details on NLSA as used in this paper, including formulas for the kernel, can be found in Text S1.

Unlike conventional EOF approaches, NLSA has rigorous connections with the spectral theory of dynamical systems \citep{Giannakis17,DasGiannakis17}, and as a result can simultaneously capture multiple intrinsic dynamical timescales from the input data without requiring ad hoc preprocessing. In particular, it can be shown that as the length of the delay embedding window increases, NLSA increasingly captures frequencies in the point spectrum of the dynamical system generating the data, filtering out the frequencies associated with the continuous spectrum. This is important since the component of the signal variability in the point spectrum exhibits periodic or quasiperiodic variability, whereas that in the continuous spectrum has the character of a stochastic or noisy source. Thus, NLSA can be thought of as a filter tailored to the underlying dynamical system, which does not eliminate signal components with respect to a particular timescale, but rather removes the components with respect to lack of temporal coherence. This property is useful in climate dynamics applications where one is interested in objective identification of phenomena spanning multiple timescales, as well as characterization of their modulating relationships (e.g., the annual cycle, ENSO, and their associated combination modes). It should be noted that even though the skill of NLSA in capturing discrete spectral components is maximal in the theoretical limit of infinitely many lags, the number of lags is practically limited by considerations such as the timespan of the available data. \citet{BushukEtAl14} extended the NLSA algorithm described above to multivariate datasets through the use of product kernels from different measured quantities. 
%It has been observed that these interactions have some characteristics of interest, such as their finer spatial patterns and regional impact.

As in SG, we  employ a class of anisotropic kernels called cone kernels \citep{Giannakis15},  and perform coupled NLSA in some of the HadISST analyses. In addition, we use a 12-year embedding window in the analyses of CCSM4 data and a 4-year embedding window for the HadISST data. The shorter embedding window compared to the CCSM4 analyses is due to the shorter timespan of the HadISST data.  The high temporal coherence and clear timescale separation in our recovered modes, which would be difficult to achieve via classical methods, rely on both the long embedding window employed and the properties of cone kernels. As a sensitivity analysis, we have verified the robustness of our results against the length $q$ of the embedding window and the cone kernel parameters (see Text~S1). In order to clearly resolve all of the ACW modes presented in Section~\ref{secRes},  $\gtrsim 6$ yr ($q \gtrsim 72$) embedding windows are generally required in CCSM4 Antarctic data and  $\gtrsim 4$ yr   ($q \gtrsim 48$) in HadISST data. Shorter,  $\gtrsim 4 $ yr ($ q \gtrsim 48$), embedding windows are enough to capture robust ENSO modes from the Indo-Pacific SST data and wavenumber-2 ACW modes from Antarctic data, as well as the associated leading combination modes in both CCSM4 and HadISST. As expected theoretically, longer embedding windows improve the timescale separation capability of NLSA and  its ability to capture low-variance modes such as decadal modes and higher-order combination modes. As with any data analysis technique, the quality of the recovered NLSA modes depends on the timespan of the available data. In general, the method's data requirements are comparable than SSA \citep{SlawinskaGiannakis17}. As part of our sensitivity analysis, we have verified that the NLSA modes recovered from 40 yr portions of the CCSM4 data are of comparable quality to their HadISST counterparts. In addition, our results are qualitatively robust against changes of the cone kernel parameter values (see Text~S1).

\section{\label{secRes}Results}

Applied to the datasets described in Section~\ref{secData}, NLSA yields a hierarchy of modes that can be grouped into (1) periodic modes associated with the annual cycle and its harmonics; (2) interannual modes and their associated modulations by the periodic modes, analogous to ENSO combination modes  \citep{McGregorEtAl12, StueckerEtAl13,StueckerEtAl15}; (3) decadal modes;  
(4) trend-like modes. In this work, we focus on the family of modes (2). For notational simplicity, we label the eigenfunctions $ \phi_k $ using the indices $ k $ determined from their ordering within the family (2) as opposed to the full NLSA spectrum.

\subsection{\label{secWave2}Wavenumber-2 ACW modes}

\begin{figure}
\centering
\includegraphics[width=\linewidth]{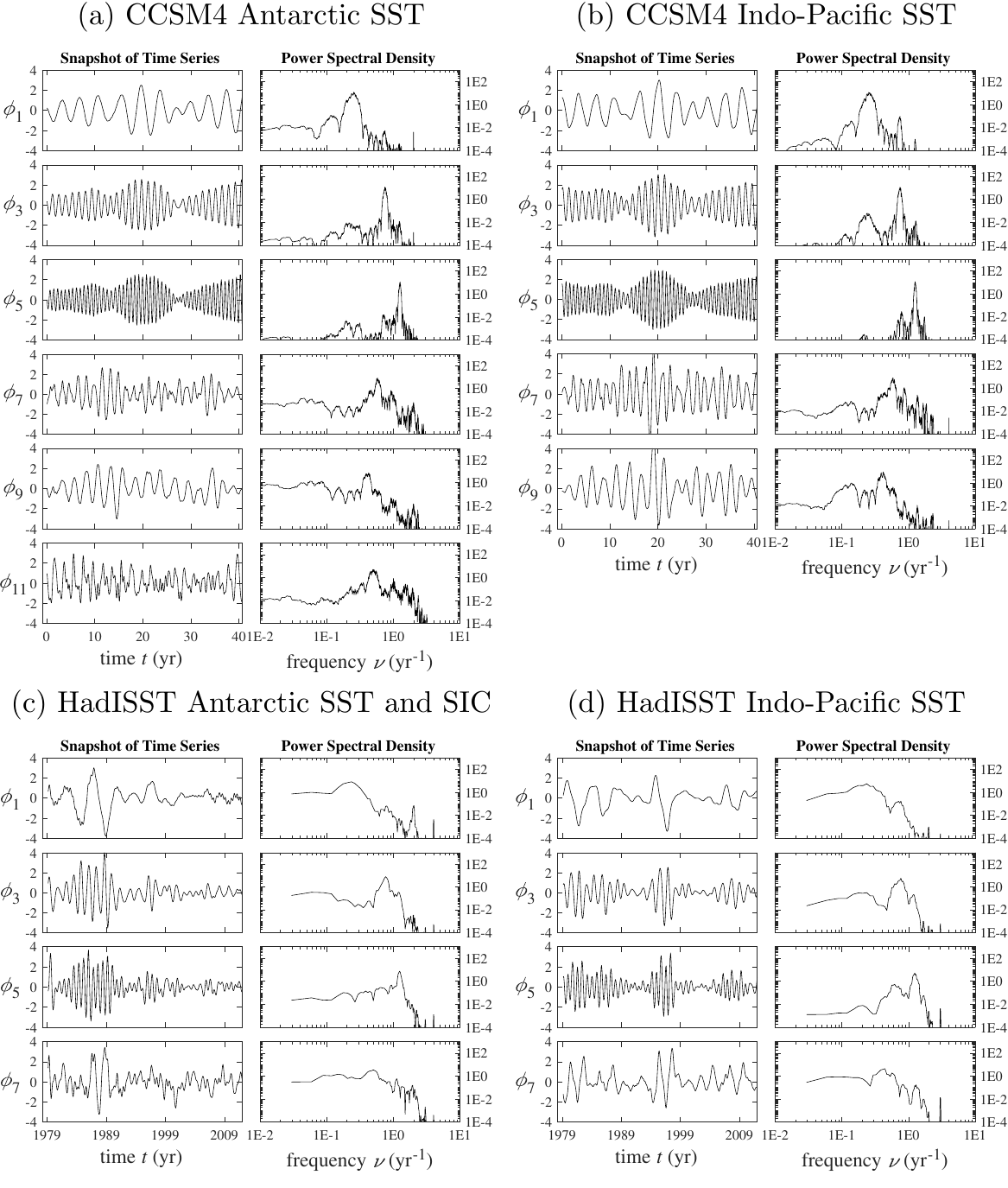}
\caption{Eigenfunction time series and power spectral densities of representative NLSA modes recovered from CCSM4 SST (a, b) and HadISST SST and SIC (c, d) data. In (a, b), only 40 yr portions of the 1300 yr timespan of the CCSM4 dataset are shown. Modes $ \phi_1 $ are primary wavenumber-2 ACW (a, c) and ENSO (b, d) modes. Modes $ \phi_3 $ and $ \phi_5 $ are combination modes with $ (m, n ) = ( 1, -1 ) $ and $ ( 1, 1) $ respectively. Modes $ \phi_7 $ and $ \phi_9 $ are higher-order combination modes with $ ( m, n ) = ( 2, 0 ) $ and $ (2, -1 ) $ respectively. Mode $ \phi_{11} $ represents the wavenumber-3 ACW. The power spectral densities were estimated via the multitaper method \citep[][]{GhilEtAl02,Thomson82}.}
  \label{figure1}
\end{figure}

\begin{figure}
  \includegraphics[width=\linewidth]{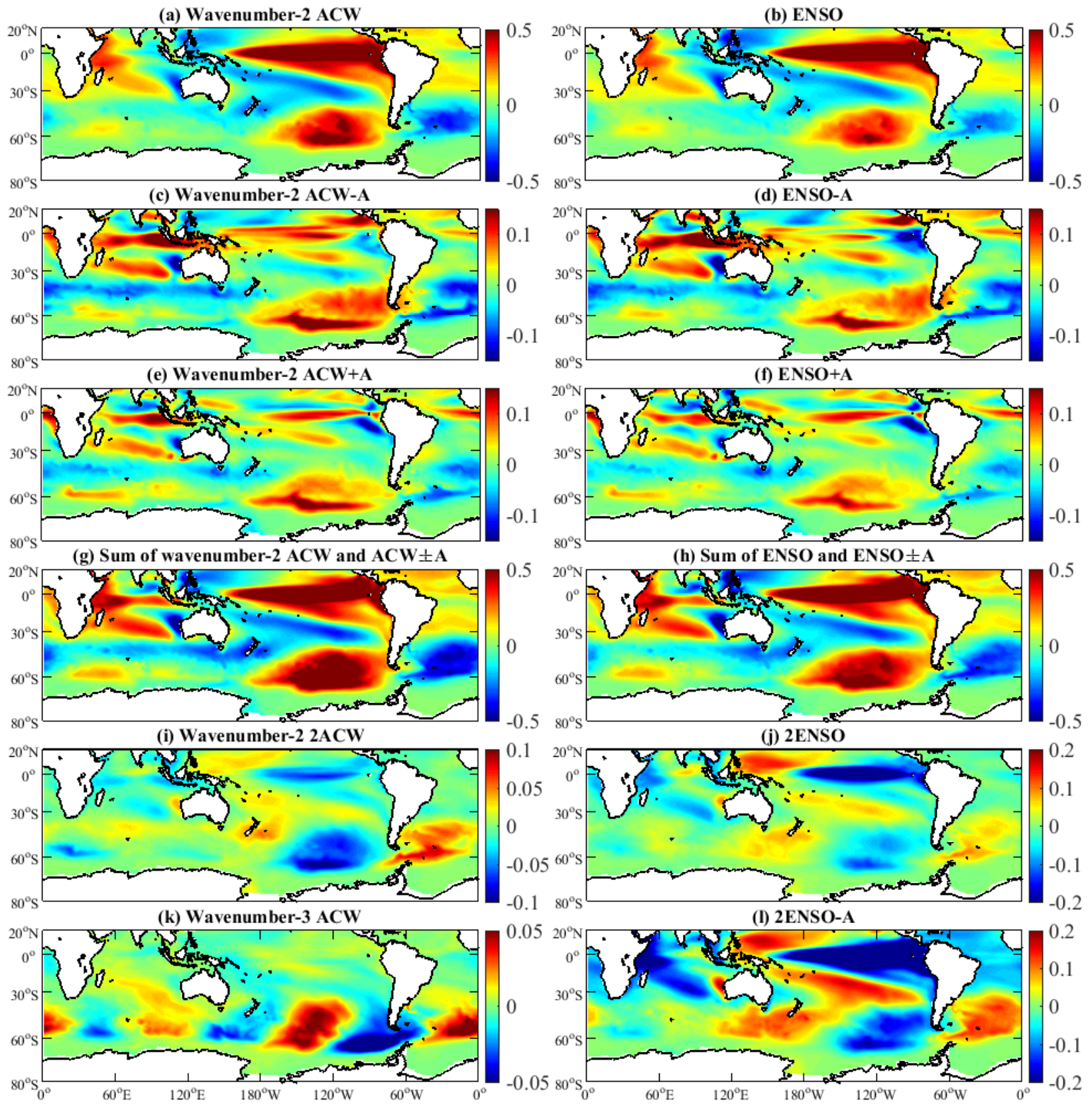}
  \caption{Snapshots of reconstructed SST anomalies (in K) based on NLSA modes recovered from CCSM4 input data. Left-hand column: Antarctic SST input as in Figure~\ref{figure1}(a). Right-hand column: Indo-Pacific SST input as in Figure~\ref{figure1}(b).}
  \label{figure2}
\end{figure}

We begin by examining NLSA modes recovered from CCSM4 Antarctic SST data. Representative eigenfunction time series and reconstructed SST snapshots associated with the SST-derived modes are shown in Figures~\ref{figure1}(a) and~\ref{figure2}, respectively. The dynamic evolution of the patterns in Figure~\ref{figure2} is visualized in Animations~S1. 

As shown in Figure~\ref{figure1}(a), the leading two NLSA modes derived from Antarctic SST, $ \{ \phi_1, \phi_2 \} $, form an oscillatory ($90^\circ$ out-of-phase) pair with a peak frequency  $f_{ACW} \approx 0.25 $~yr$^{-1}$ characteristic of the ACW and a low-frequency (decadal) amplitude envelope.  This peak frequency is in good agreement with the corresponding ACW frequency identified in the observational studies of \citet{Cerrone2017a, Cerrone2017b} (SST and SLP data), and is also close to the 5 yr$^{-1}$ frequency identified by \cite{V2003}  (SST, SIC, and multivariate tropospheric data). In spatiotemporal reconstructions (Figure~\ref{figure2}(a) and Animation~S1), these modes give rise to a wavenumber-2, eastward-propagating ACW signal, which attains its maximum strength in the eastern Pacific sector of the Southern Ocean. As a cluster of anomalies associated with this signal crosses the Drake Passage, it weakens in amplitude (but does not diminish completely), and regains its strength upon entering the western Atlantic sector. By that time, the opposite-sign anomaly cluster trailing it has reached the eastern Pacific sector and gained strength. This results in the appearance of a transient dipole anomaly pattern between the eastern Pacific and western Atlantic sectors. The large anomaly amplitudes occurring during this phase in conjunction with the fact that EOF analysis optimizes for variance, are a likely explanation of the origin of a quasi-stationary Antarctic dipole pattern in the dominant mode recovered by that technique \citep{YM2001}. However, our analysis suggests that this dipole is part of a continuous eastward-propagating traveling mode with a time-dependent amplitude. Subsequently in the ACW lifecycle, the SST anomalies travel eastward into the Indian sector where they dissipate considerably, taking approximately 8 yr to circumnavigate Antarctica. %{\color{red}The spatial patterns of our wavenumber-2 ACW modes remarkably resemble those observed by \citet{V2003} and \citet{Cerrone2017b} in which the eastward-propagating SST anomalies cover a continuous path and are reinforced in the western Pacific Ocean around $60^{\circ}$S.}

\begin{figure}
  \centering
  \includegraphics[width=\linewidth]{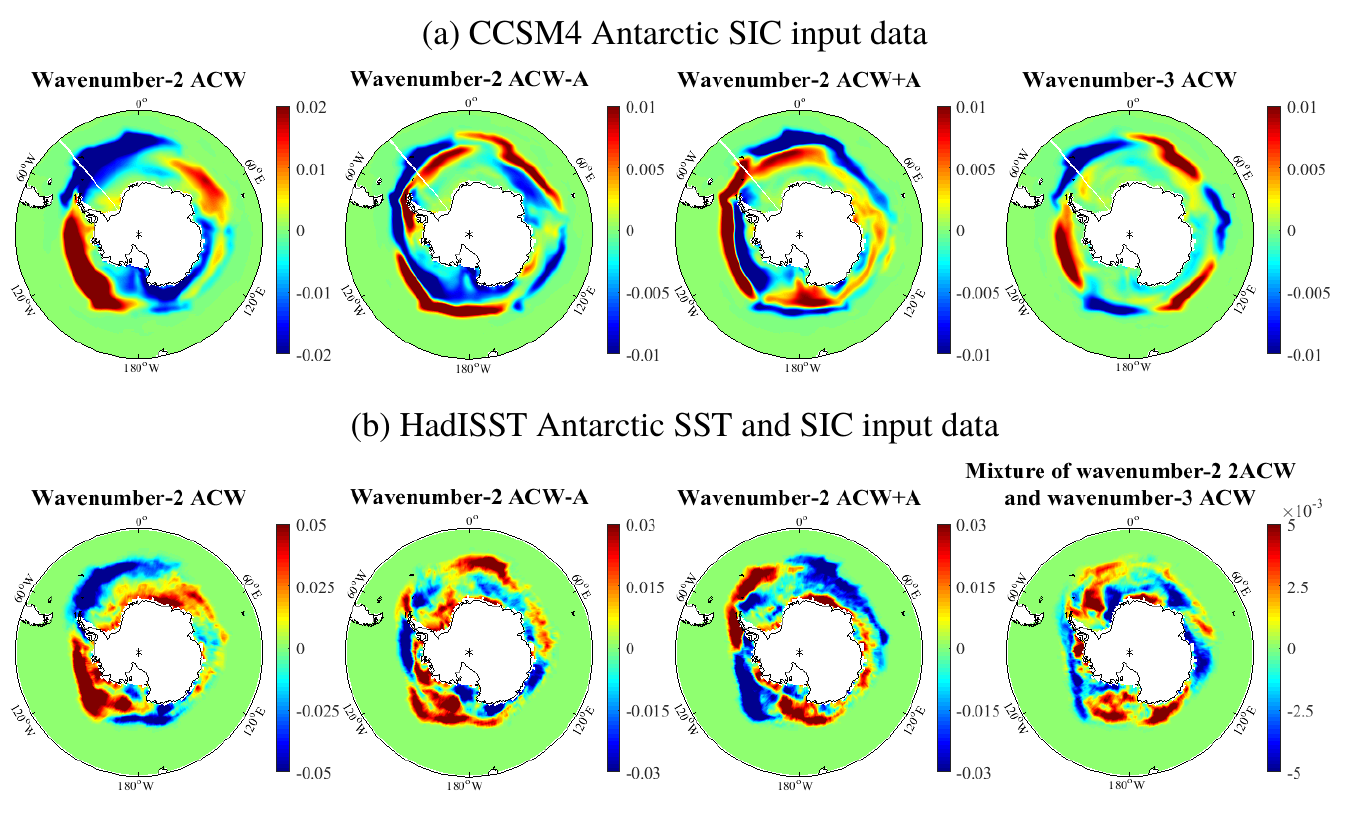}
  \caption{Snapshots of reconstructed SIC anomalies (dimensionless) based on NLSA modes recovered from (a) Antarctic SIC input data from CCSM4, (b) Antarctic SST and SIC input data from HadISST.}
\label{figure3}
\end{figure}

Meanwhile, in the tropical and subtropical latitudes, the SST anomaly patterns associated with $ \{ \phi_1, \phi_2 \} $ have a triangular ENSO-like structure. Upon closer inspection, the SST anomalies in the Southern Ocean can be traced to anomalies originating in the western subtropical South Pacific Ocean and moving southward into the Southern Ocean where they travel eastward with the Antarctic circumpolar current (as proposed by \citet{WP1998}). The teleconnections between ACW and ENSO will be further discussed in Section~\ref{secENSO}. A consistent, 0.25 yr$^{-1}$, eastward-propagating ACW2 structure is also captured by the $ \{ \phi_1, \phi_2 \} $ modes recovered from SIC data (Figure~\ref{figure3}(a) and Animation~S3). Note that the frequency of these modes differs from the dominant SIC signal recovered by \citet{Cerrone2017b} from observational data, which has a 2.7 yr periodicity.

In addition to the primary wavenumber-2 ACW modes described above, the NLSA spectrum contains a family of combination modes with peak frequencies of the form $ f_{m,n} = | m f_{ACW} + n f_{A} | $, where $ m, n $ are integers, and $ f_A = 1 $ yr$^{-1} $ is the annual cycle frequency. In Figure~\ref{figure1}(a), we show the time series of SST-derived eigenfunctions $ \{ \phi_3, \phi_4 \} $ with $ ( m, n ) = ( 1, -1 ) $ (denoted ACW$-$A), $ \{ \phi_5, \phi_6 \} $ with $ ( m, n ) =  (1, 1 ) $ (denoted ACW$+$A),  $ \{ \phi_7, \phi_8 \} $ with $( m,n ) = ( 2, 0) $ (denoted 2ACW), and $ \{ \phi_9, \phi_{10} \} $ with $ ( m, n ) = ( 2, -1 ) $ (denoted 2ACW$-$A). The peak frequencies of these modes identified from their frequency spectra are respectively $ f_{1,-1} = 0.75 $~yr$^{-1}$, $ f_{1,1} = 1.25 $~yr$^{-1} $, $ f_{2,0} = 0.58 $~yr$^{-1}$, and $ f_{2,-1} = 0.42 $~yr$^{-1}$. To examine the temporal characteristics of these modes in more detail, we consider the relationships between the complex numbers $ z_{1,0} = \phi_1 + i \phi_2 $, $ z_{1,-1} = \phi_3 + i \phi_4 $, $ z_{1,1} = \phi_5 + i \phi_6 $,  $ z_{2,0} = \phi_7 + i \phi_8 $, and $ z_{2,-1} =\phi_9 + i \phi_{10} $. As shown in Figure~S1, the quotients $ z_{1,0}/z_{1,\mp 1} $ exhibit clear spectral peaks at $ \pm 1 $ yr$^{-1} $, as expected for an annual sinusoidal signal modulated by the ACW signal $ z_{1,0} $ (see Text~S2). Similarly,  $ z_{1,0}^2 / z_{2,0}  $ and $ z_{1,0}^2 / z_{2,-1} $ exhibit spectral peaks at 0~yr$^{-1}$ and 1~yr$^{-1} $, respectively, though these peaks are not as strong as in the case of the ACW$\pm$A modes. As with the primary ACW modes, ACW combination modes can be recovered from SIC data as well.  Other combination modes, including semiannual ($ n = 2 $) and triennial ($ n = 3 $) combination modes are also resolved but not shown here. To our knowledge, it is the first time that this multiscale hierarchy of combination  modes, with a close agreement between theoretically expected and actual frequencies, has been derived from Antarctic data via objective eigendecomposition techniques.

We now turn to the spatiotemporal characteristics of the ACW combination modes in the Southern Ocean. In particular, as is evident from Animations~S1 and~S3 (especially in the Pacific and Atlantic sectors of the Southern Ocean), and consistent with the fact that the phase velocities of $ z_{1,0} $ and $ z_{1,\mp1} $ have opposite signs, the ACW$-$A and ACW$+$A modes have opposite direction of propagation, with the former traveling westward and the latter eastward. Moreover, in the SIC field, these modes exhibit appreciable wavenumber-2 variability in the meridional direction (Figure~\ref{figure3}(a)), consistent with the seasonal growth and melting of Antarctic sea ice in response to the seasonal cycle. Analogous relationships between annual-cycle modulated patterns and sea-ice reemergence phenomena \citep{BlanchardEtAl11,HB2013} have also been identified in the Arctic \citep{BushukEtAl14,BushukGiannakis17}, but have not received similar attention in the Southern Ocean. Another notable feature of the ACW$\pm$A modes is that they carry more variance than the primary ACW modes over the Indian Ocean sector ($20^{\circ}$E--$170^{\circ}$E), the regions near $70^{\circ}$S in the Pacific sector,  and those near $50^{\circ}$S in the Atlantic sector (the latter, in SIC only). Figure 4 displays the difference between the explained variances of the SST and SIC anomalies reconstructed from combination modes and those from the primary ACW modes. These patterns, which are consistent with the meridional structures in the spatial patterns in Figure~\ref{figure3}, imply that the seasonal cycle can transport ACW signals to lower and higher latitudes in the the Southern Ocean, and play a more significant role in the Indian sector where the activity of the primary ACW modes is diminished. Overall, due to the meridional variability and opposite sense of propagation of the ACW$\pm$A modes, the reconstructed SST and SIC anomalies based on them and the primary ACW modes (Figure~\ref{figure2}(g) and Animations~S1 and~S3), acquire a meridional oscillation, as well as an intricate zonal propagation structure, particularly over the eastern Indian Ocean sector where a standing wave component is observed in the total SIC anomaly field. Our analysis allows one to attribute these behaviors to combination modes between the primary wavenumber-2 ACW and the annual cycle.

\begin{figure}
  \centering
  \includegraphics[width=0.7\linewidth]{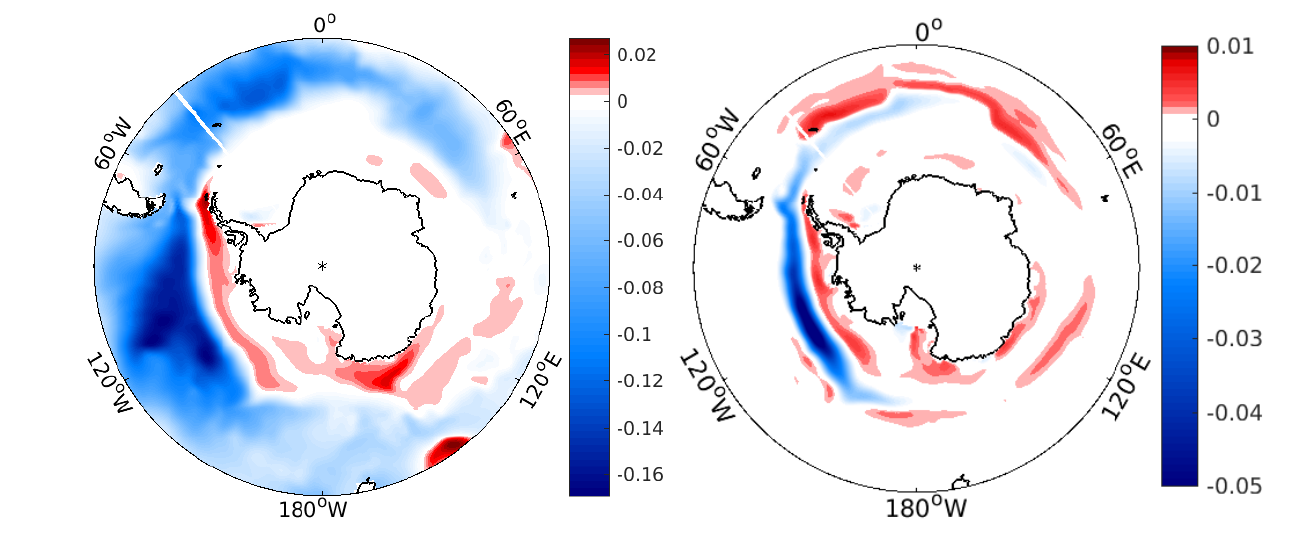}
  \caption{Difference between the SST (left) and SIC (right) explained variance due to the ACW$\pm$A and primary ACW modes. Explained variance values are computed at each gridpoint relative to the climatological variance after removal of the seasonal cycle.}
\label{figure4}
\end{figure}

The picture described above can be further refined by including additional ACW combination modes. For instance, Figure~\ref{figure2}(i) shows the reconstruction of the quadratic mode 2ACW, in which SST anomalies show a different structure from the primary ACW mode due to the shorter period and stronger intermittency of that mode. Another example is the semiannual modulation at the frequencies $ f_{1,\pm2} $ (not shown), where westward/eastward propagation and a strong signal in the Indian sector can be observed as in the $ f_{1,\pm 1} $ modes. In summary, at any given time,  the total wavenumber-2 ACW state will be weakened or reinforced by the annual-cycle combination modes and higher-order harmonics in the $ f_{m,n} $ frequency spectrum, which also influence its spatial structures and propagation characteristics.

Next, we discuss analogous analyses performed on HadISST Antarctic data for the satellite era with a shorter, 4 yr, embedding window and show that most of the conclusions drawn from the CCSM4 analyses also hold in HadISST, despite the smaller number of samples and observational biases and noise. In this case, higher-quality modes were obtained from coupled NLSA applied to SST and SIC data; in what follows, we restrict attention to results obtained via this approach.

\begin{figure}
  \includegraphics[width=\linewidth]{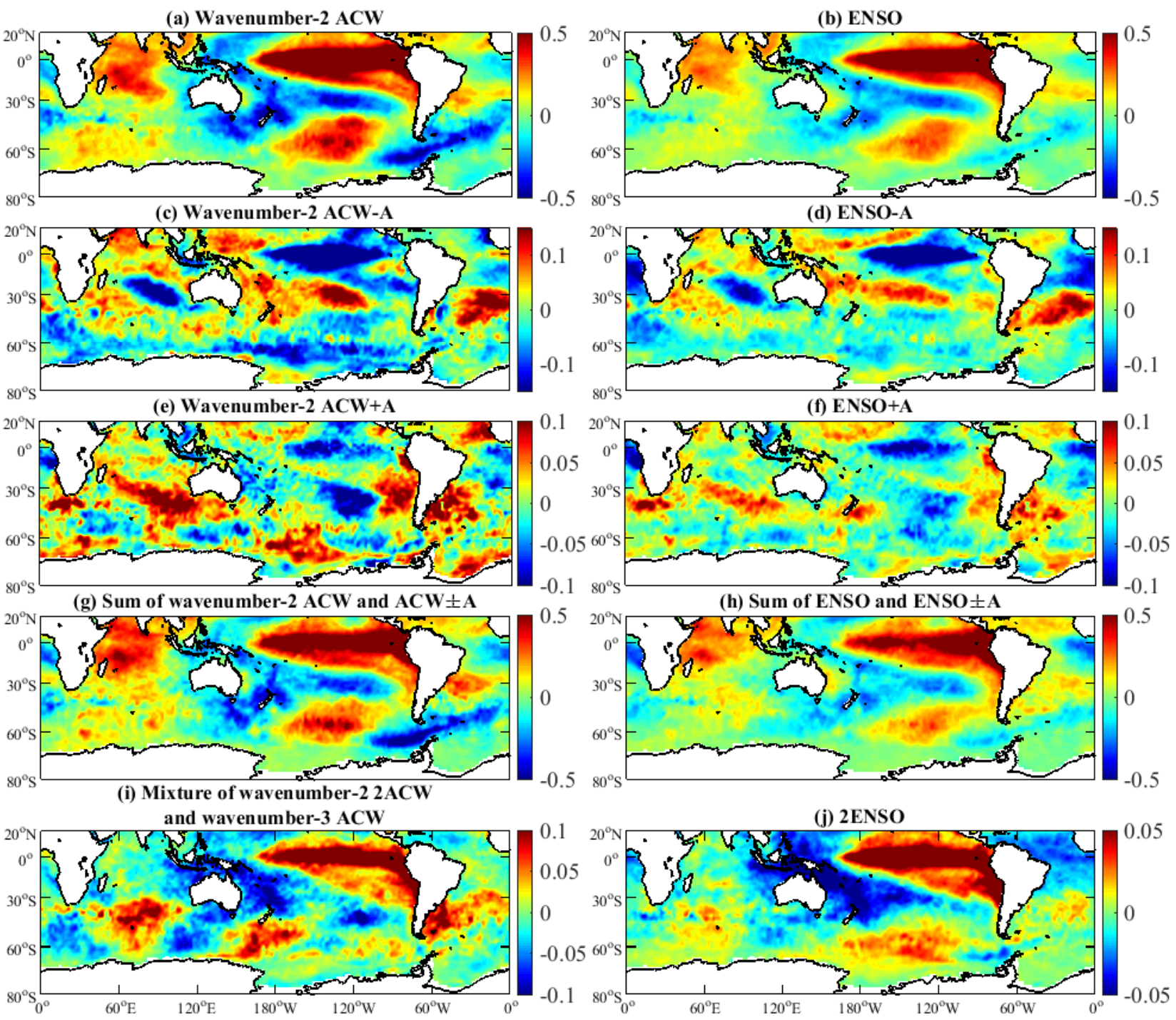}
  \caption{As in Figure~\ref{figure2}, but for NLSA modes derived from satellite-era HadISST input data. Left-hand column: Antarctic SST and SIC input as in Figure~\ref{figure1}(c). Right-hand column: Indo-Pacific SST input as in Figure~\ref{figure1}(d).}
  \label{figure5}
\end{figure}

Figure~\ref{figure1}(c) shows representative eigenfunctions associated with the wavenumber-2 ACW and its combination modes with the annual cycle. Evidently, the eigenfunction time series in Figure~\ref{figure1}(c) are noisier than their CCSM4 counterparts in Figure~\ref{figure1}(a), but remain fairly consistent with the frequency cascade $ f_{m,n} = | m f_{ACW} + n f_A | $. Specifically, the primary ACW frequency peak of the HadISST-derived eigenfunctions $ \{ \phi_1, \phi_2 \} $ is $ f_{ACW} = 0.24 $ yr$^{-1}$ (again consistent with those in \citet{V2003} and \citet{Cerrone2017a,Cerrone2017b}), and the corresponding peaks of $ \{ \phi_3, \phi_4 \} $ (0.77 yr$^{-1}$) and $ \{ \phi_5, \phi_6 \} $ (1.24 yr$^{-1}$) are in good agreement with the theoretical combination frequencies with $ ( m, n ) = ( 1, -1 ) $ and $ ( 1, 1  ) $, respectively. The peak frequency of $ \{ \phi_7, \phi_8 \} $ (0.47 yr$^{-1}$) is close to the $  ( 2, 0 ) $ combination frequency, but the time series of these modes are markedly noisier than their $ m = 1 $ counterparts. In addition to identifying spectral peaks, the association of specific $ (m, n ) $ values with the eigenfunctions in Figure~\ref{figure1}(c) was corroborated with spectral tests involving the complex numbers $ z_{m,n} $ as described above for CCSM4 data (see Figure S2). The corresponding spatiotemporal patterns (Figures~\ref{figure5} and~\ref{figure3}(b), Animations~S4 and~S6) also exhibit broadly consistent features with those of the CCSM4-derived wavenumber-2 ACW modes, including in particular the sense of propagation and seasonal meridional pulsation of the ACW$\pm$A modes. The latter is best visualized in the SIC reconstructions in Figure~\ref{figure3}(b) and Animation~S6, where the ACW$-$A and ACW$+$A modes exhibit a clear sense of westward and eastward propagation in the Pacific and Atlantic sectors of the Southern Ocean, respectively. 

\subsection{\label{secWave3}Wavenumber-3 ACW modes}

Another mode recovered by NLSA from Antarctic SST and SIC data in CCSM4 is a wavenumber-3 ACW mode. In the case of the SST-derived modes, the primary wavenumber-3 ACW structure (denoted ACW3) is represented by eigenfunctions $ \{ \phi_{11}, \phi_{12} \} $ (Figure~\ref{figure1}(a)), which form an oscillatory pair as the wavenumber-2 ACW modes. The peak frequency of the $ \{ \phi_{11}, \phi_{12} \} $ pair is  $ f_{ACW3} = 0.49 $ yr$^{-1}$. Our ACW3 modes have a somewhat shorter period than the 3 and 2.7 yr periods identified by \citet{V2003} and \citet{Cerrone2017b}, respectively. As described below, a possible reason for this discrepancy is that the characteristic  ACW3 frequency is close to the frequencies of a number of higher-order ACW2 combination modes. In the spatiotemporal domain (Figure~\ref{figure2}(k) and Animation~S1), this pair exhibits a prominent eastward-propagating wavenumber-3 structure with weak activity over the tropics and subtropics, in agreement with the patterns in \citet{Cerrone2017b}. The wavenumber-3 ACW pair derived from SIC data, shown in Figure~\ref{figure3}(a) and Animation~S3, is largely consistent with the SST-derived pair. 

On the other hand, a distinct wavenumber-3 ACW mode is absent in the spectrum of HadISST-derived NLSA modes from Antarctic SST and SIC data. Upon closer inspection, it can be seen that the $ \{ \phi_7, \phi_8 \} $ pair in Figure~\ref{figure1}(c) exhibits a wavenumber-3 pattern in the Southern Ocean, particularly over the Indian Ocean longitudes as shown in Figure~\ref{figure5}(i), but that pattern is rather weak and intermittent.

The close proximity of $ f_{ACW3} $ with the frequencies of the quadratic wavenumber-2 ACW modes (e.g., $ f_{2,0} $ and $ f_{2,-1} $) makes the detection of this pattern as an individual mode particularly challenging with objective eigendecomposition techniques. Indeed, some mixing between the wavenumber-3 and the wavenumber-2, $ m = 2 $, ACW combination modes (e.g., Figure~\ref{figure5}(i)) can be seen as the former sometimes develops an ENSO-like pattern over the tropics, and the latter exhibits a weak wavenumber-3 structure in the Southern Ocean. This behavior also takes place, though at a significantly lesser extent, in the CCSM4-derived modes (see Animation~S1). The detectability of pure wavenumber-3 ACW modes is further compounded by their low amplitudes (e.g., in Figure~\ref{figure2}, an order of magnitude smaller compared to the wavenumber-2 ACW), making them harder to recover from shorter datasets using shorter delay-embedding windows. Due to these reasons, the absence of a distinct wavenumber-3 ACW mode in our HadISST analysis does not imply the absence of this mode in nature. Similar mixing also occurs between the quadratic mode $(m,n)=(2,0)$ and its modulation $(m,n)=(2,-1)$ due to their similar frequencies, making the westward propagation of the latter mode unclear in this case. A co-occurrence of PSA and zonal wavenumber 3 patterns has also been observed in the studies of \citet{Cerrone2017a,Cerrone2017b}, who find that the spectrally dominant components of ACWs can change from wavenumber 3 to wavenumber 2 during eastward propagation \citep{Cerrone2017b}, and the influence of wavenumber-3 patterns on SST anomalies are significantly weaker than those associated with the PSA \citep{Cerrone2017a}.

\subsection{\label{secENSO}ENSO teleconnections}

Given the spatiotemporal characteristics of the Antarctic-derived modes described in Section~\ref{secWave2}, it is natural to study in more detail their connections with tropical Indo-Pacific variability, and in particular ENSO. Representative eigenfunctions obtained via NLSA from CCSM4 Indo-Pacific SST data are shown in Figure~\ref{figure1}(b). As is visually evident (and quantitatively verified via frequency-domain tests in Figure~S1), the temporal structure of the eigenfunctions bears strong similarities with the eigenfunctions in Figure~\ref{figure1}(a) derived from Antarctic data. In particular, the NLSA spectrum from Indo-Pacific SST data features a leading pair of interannual modes, $ \{ \phi_1, \phi_2 \} $, with a broad, yet prominent spectral peak at $ f_{ENSO} = 0.25 $ yr$^{-1} $ and an associated family of combination modes at the frequencies $ f_{m,n} = | m f_{ENSO} + n f_A | $. Other modes in Figure~\ref{figure1}(b) are ENSO combination modes $ \{\phi_3,\phi_4\} $ with $ ( m, n ) = ( 1, -1 ) $ and peak frequency $ f_{1,-1} = 0.74 $ yr$^{-1}$,  combination modes $ \{\phi_5,\phi_6\} $ with $ ( m, n ) = ( 1, 1 ) $ and $ f_{1,1} = 1.25 $ yr$^{-1} $, quadratic ENSO modes $ \{\phi_7,\phi_8\} $ with $ ( m, n ) = ( 2, 0) $ and $ f_{2,0} = 0.58$ yr$^{-1}$, and combination modes $ \{ \phi_9,  \phi_{10} \} $ with $ ( m, n ) = ( 2, -1 ) $ and $f_{2,-1} = 0.41$ yr$^{-1}$. For the sake of convenience, we denote these modes by ENSO$\pm$A, 2ENSO and 2ENSO$-$A, respectively. Snapshots and videos of the corresponding reconstructed SST anomalies are displayed in Figure~\ref{figure2} and Animation~S2.

As suggested by the frequencies listed above, there exist strong linkages between the ENSO modes derived from Indo-Pacific SST data and the wavenumber-2 ACW modes derived from Antarctic SST data. In particular, the correlation coefficients between the ENSO and ACW eigenfunction time series, as well as the corresponding ENSO$\pm$A and ACW$\pm$A combination modes, are all $ \simeq 0.9 $, while those between the quadratic ENSO and ACW modes are $\simeq 0.45$. Likewise, as can be seen by comparing the left- and right-hand columns in Figure~\ref{figure2}, the spatial patterns of reconstructed SST anomalies associated with the Antarctic-derived modes resemble considerably those obtained from the Indo-Pacific-derived modes. That is, the Indo-Pacific-derived primary ENSO modes project strongly to a clear, eastward-propagating, wavenumber-2 pattern in the Southern Ocean that resembles closely the structure of the leading ACW mode in that region, and the ENSO$\pm$A modes give rise to westward/eastward propagating and meridionally pulsating anomalies similar to those associated with the ACW$\pm$A modes. A similarly good agreement between the Indo-Pacific- and Antarctic-derived patterns is also observed in the tropics and subtropics.

Combination modes between interannual modes and the annual cycle were previously identified by \cite{McGregorEtAl12} and \citet{StueckerEtAl13, StueckerEtAl15} in a series of studies on ENSO variability. They found that the combination frequencies between ENSO and the annual cycle (in our notation, $ f_{1,\pm 1} $) are a nonlinear outcome of the atmosphere-ocean dynamical coupling in response to the annually varying solar insolation, resulting in a southward shift of tropical zonal winds that plays an important role in explaining ENSO seasonality and termination. Our results show that analogous combination  modes can also be identified in the context of ACW variability, providing a framework for studying physical processes associated with the coupled ocean-atmosphere-ice dynamics of the ACW. In particular, it is known that nonlinear interactions with cyclones and other synoptic systems (e.g., modulation of the spatial density and intensity of cyclones by SST gradients) play an important role in ACW dynamics \citep{W2006}, and even though these phenomena are not directly represented in our monthly-averaged data, they may play an important role in the emergent behavior captured by the NLSA modes. Note that the combination modes recovered by EOF analysis \citep{StueckerEtAl13} actually exhibit a mixture of the theoretical ENSO$\pm$A frequencies, but NLSA clearly separates these frequencies and their associated spatial patterns, while also recovering higher-order combination modes.

\begin{figure}
\centering
\includegraphics[width=.55\linewidth]{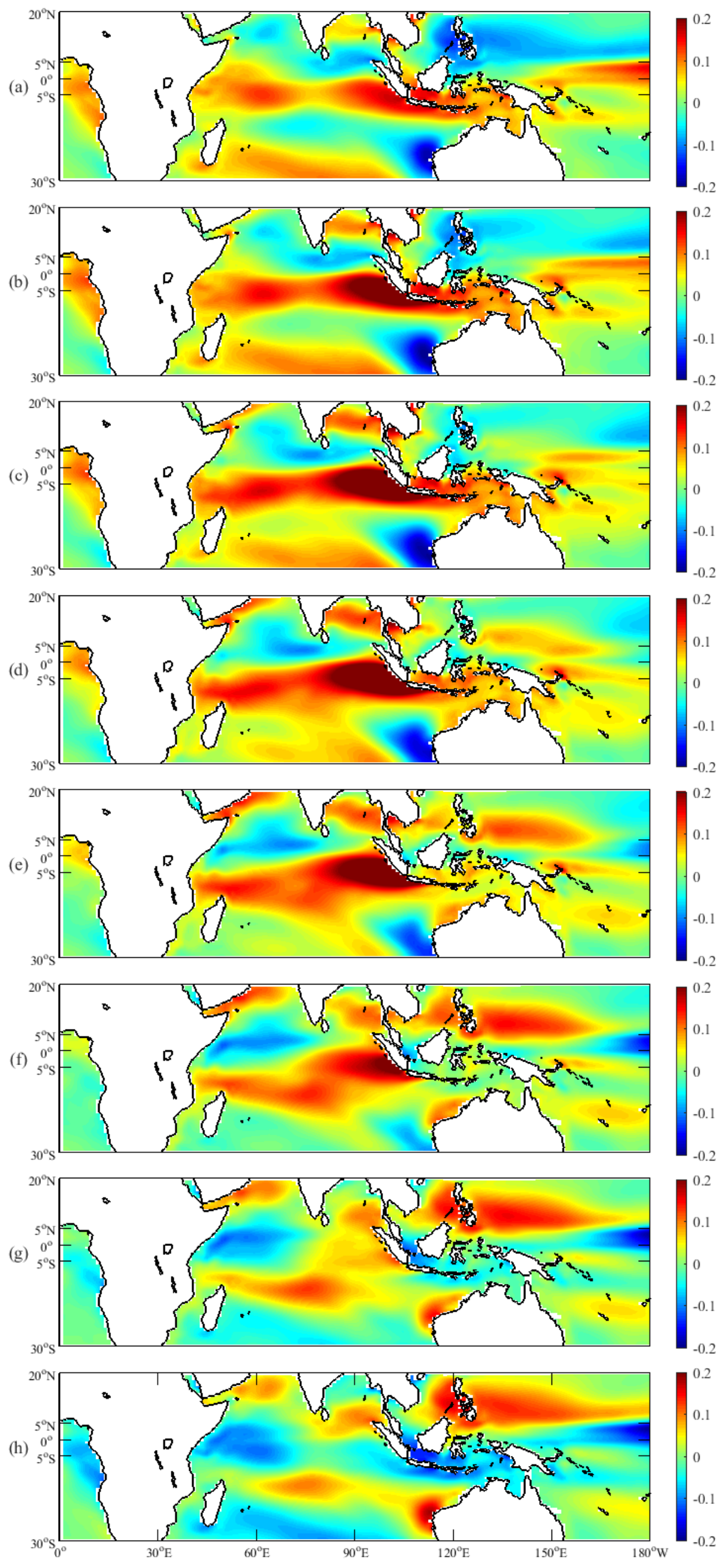}
\caption{Snapshots of reconstructed SST anomalies (in K) from the ENSO$-$A mode derived from CCSM4 Indo-Pacific SST input data for eight consecutive months (i.e., about half ENSO$-$A period) from December of simulation year 22 (a) to July of simulation year 23 (h). Dipolar anomaly patterns in the tropical Indian Ocean can be clearly observed in (e--h).}
\label{figure6}
\end{figure}

Another notable feature of the ENSO$\pm$A modes is their prominence in the tropical Indian Ocean regions associated with the Indian Ocean dipole (IOD) \citep{SajiEtAl99,WebsterEtAl99}. For instance, a clear dipole is observed in certain phases during the lifecycle of ENSO$-$A mode, with positive SST anomalies developing over the tropical Indian Ocean around $5^{\circ}$S, propagating southeastward, and forming a dipole with a cluster of negative anomalies which emerges around $(5^\circ\mathrm{N}, 90^\circ\mathrm{E})$ and propagates westward to east Africa (see Figure~\ref{figure6}). This pattern repeats in a sign-reversed fashion in the ensuing months, and similar dipoles are also observed in ENSO$+$A reconstructions with a different periodicity. This connection between ENSO combination modes and the IOD has been noted in previous studies using NLSA-derived ENSO combination modes from Indo-Pacific SST data (SG), or theoretically-derived combination modes \citep{StueckerEtAl17}. SG, in particular, noted a number of propagating patterns in the tropics and mid-latitudes of the Indian Ocean basin associated with these modes, undergoing multiple reflections at the western (Africa) and eastern (Australia and Maritime Continent) coasts of the basin. Intriguingly, we have shown here that the same combination modes, extracted from either Antarctic or Indo-Pacific data, are also associated with prominent traveling anomaly patterns at polar latitudes of the Indian Ocean sector, extending all the way to the coast of Antarctica. As in the case of the ACW$\pm$A combination modes (see Section~\ref{secWave2}), the ENSO$\pm$A modes carry more SST variance in the Indian Ocean, especially in the eastern equatorial part of the basin where the IOD is active. In particular, in the region $90^{\circ}$E--$110^{\circ}$E, $10^{\circ}S$--$10^{\circ}$N approximately $11\%$ variance is explained by the ENSO combination modes, whereas only $\sim 2\%$ is carried by the primary ENSO modes.

\begin{figure}
\centering
\includegraphics[width=\linewidth]{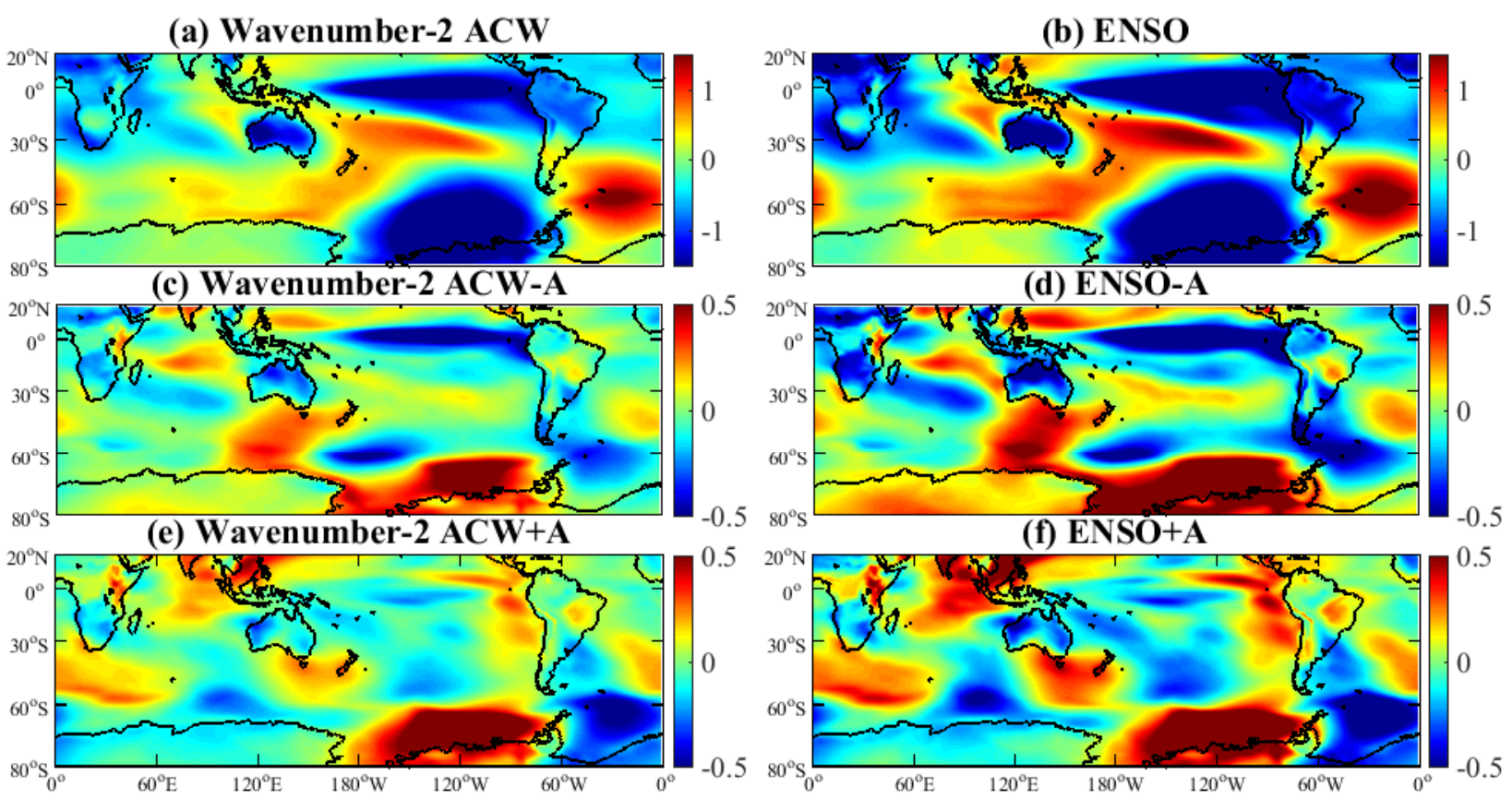}
\caption{Snapshots of reconstructed 850hPa geopotential height anomalies (in m) based on the wavenumber-2 ACW (left-hand column) and ENSO (right-hand column) modes and their associated combination modes derived from Antarctic and Indo-Pacific SST data from CCSM4, respectively, during a strong La Ni\~na event.}
\label{figure7}
\end{figure}

To further examine the ENSO-ACW relationship described above, we have reconstructed the 850hPa geopotential height anomalies associated with the ENSO and ACW modes from CCSM4. These reconstructions from both the primary modes and the corresponding combination modes display clear wavetrain patterns in mid- to high latitudes emanating southeastward from Australia to the southeastern Pacific ocean (Figure~\ref{figure7}), resembling the Pacific-South American (PSA) pattern \citep{K1988, MH1998, BC2001,I2016}. A number of studies have found relationships between the PSA and both ENSO \citep{K1989, M2000, MP2001, I2016} and the ACW \citep{C1998, BC2001,Cerrone2017a, Cerrone2017b, CF2018}. Our results confirm the close connection among ENSO, PSA, and ACW signals at the level of the fundamental modes, and moreover identify similar PSA-like patterns for the associated combination modes. 

While our results have established strong similarities between the families of ENSO and wavenumber-2 ACW modes, a notable difference in the Indo-Pacific analysis is the absence of a distinct wavenumber-3 ACW frequency in the eigenfunction time series (Figure~\ref{figure1}(b)), or a discernible wavenumber-3 structure in the Southern Ocean in the corresponding spatiotemporal patterns (Figure~\ref{figure2}). This suggests that the wavenumber-3 ACW has no strong link to the tropics and subtropics. In particular, the absence of this mode in the Indo-Pacific-derived mode family is consistent with the hypothesis that wavenumber-3 variability is not remotely forced by ENSO \citep{ C1998, BC1999, V2003,I2015}.

Next, we validate our results derived from CCSM4 against modes recovered from Indo-Pacific HadISST data. Eigenfunction time series for these modes are shown in Figure~\ref{figure1}(d). In this case, the correlation coefficients between the ENSO and ACW modes and the corresponding leading ($n=1$) combination modes are about 0.6, and 0.33 for quadratic ($m=2$) modes. The corresponding spatiotemporal reconstructions are displayed in Figure~\ref{figure5} and Animation~S5. As in CCSM4 analysis, the primary ENSO modes are qualitatively consistent with the primary ACW modes in that they project to wavenumber-2, eastward-propagating structures in the Southern Ocean. The ENSO$\pm$A modes are also in accord with the ACW$\pm$A modes, in the sense that they bear strong similarities in their spatial patterns and yield clear westward- and eastward-propagating anomalies in the Pacific sector of the Southern Ocean, respectively. As in CCSM4, there is no clear wavenumber-3 ACW pattern in the HadISST Indo-Pacific-derived modes.

\subsection{Decadal variability}

Although this study focuses on seasonal to interannual variability, it is worthwhile examining the linkages between the ACW and Pacific decadal variability. For instance, \citet{P2007} has found a robust link between cyclonic activity in the high latitudes of Southern Hemisphere and the Pacific Decadal Oscillation (PDO), characterized by stronger cyclonic activity during the positive PDO phase. As stated in Section~1, SG identified a decadal mode (the WPMM), which  correlates significantly with the decadal amplitude envelope of the primary ENSO modes from NLSA (see Figure~\ref{figure1}(b)). The WPMM features a prominent cluster of SST anomalies in the western tropical Pacific. Its phase with negative SST anomalies in that region (which can last for a number of decades) is associated with anomalous surface divergence, and thus anomalous decadal westerlies in the central Pacific and a flatter zonal thermocline profile. Modeling studies \citep[e.g.,][]{RodgersEtAl04} have found this background state to be favorable to increase ENSO activity, consistent with the empirical results from NLSA. 

The close correspondence between ENSO and wavenumber-2 ACW modes identified here implies the latter also exhibit a significant decadal amplitude modulation by the WPMM. Given that, a natural question to ask is whether the WPMM can be detected from Antarctic SST and/or SIC data. Here, we have attempted to identify the WPMM from Antarctic data using a variety of data sources and embedding windows, including a 20 yr embedding window as in SG, yet none of these analyses yielded a distinct decadal mode resembling the WPMM. This is consistent with the hypothesis that the main influence of this mode in the Southern Ocean is through amplitude modulations of the interannual modes, mediated through ENSO teleconnection (note that such an effect would not appear as a distinct mode, since it is already represented by the decadal envelope of the ACW modes).

\section{Concluding remarks}

In this work, we have applied NLSA to study variability in the Southern Ocean associated with the ACW without ad hoc preprocessing of the input data, and explore the covariability of ACW with ENSO. In particular, we identified a hierarchy of ACW modes that delineates the fundamental eastward-propagating wavenumber-2 and 3 ACW signals. By avoiding data prefiltering to remove trends, or isolate signals in a particular frequency band, this hierarchy also includes combination modes between the fundamental ACW patterns and the harmonics of the annual cycle that evolve on either slower or faster (up to seasonal) timescales than the fundamental 4 yr ACW timescale. These combination modes can also have either eastward or westward direction of propagation, while exhibiting meridional pulsations, producing spatially inhomogeneous propagation characteristics and displaying more complex activity in the Indian Ocean. Moreover, we showed that the wavenumber-2 ACW mode family recovered from Antarctic SST data correlates strongly with the ENSO mode family recovered from Indo-Pacific SST. In contrast, no such correspondence was identified for wavenumber-3 ACW modes, consistent with these modes being attributed to local sea-air interaction.  As future work, we plan to refine and further assess the patterns identified here via newly developed Koopman operator techniques applied to high-dimensional space-time climate data \citep{SlawinskaEtAl17}. The close correlation between the wavenumber-2 ACW and ENSO mode families may be also useful in predictive and mechanistic studies, which we plan to pursue in the future via an extension of NLSA for analog prediction \citep{ComeauEtAl17}.

%%%%%%%%%%%%%%%%%%%%%%%%%%%%%%%%%%%%%%%%%%%%%%%%%%%%%%%%%%%%%%%%
%
%  ACKNOWLEDGMENTS

%\acknowledgments
\section*{Acknowledgments}
D.\ Giannakis acknowledges support from ONR MURI Grant 25-74200-F7112,  ONR Grant N00014-14-0150, and NSF Grant DMS-1521775. X.\ Wang was supported as a PhD student from the last two grants. J.\ Slawinska received support from NSF EAGER grant 1551489. The CCSM4 data were downloaded from the Earth System Grid website (\url{http://www.earthsystemgrid.org}). The HadISST data were downloaded from the Met Office Hadley Centre website (\url{http://www.metoffice.gov.uk/hadobs/hadisst/}). We thank Mitch Bushuk, Malte Stuecker, and Xiaojun Yuan for stimulating conversations.

\section*{Supporting information}
The following supporting information is available as part of the online article:\\
{\bf Text S1.} Overview of NLSA algorithms.\\
{\bf Text S2.} Tests for combination modes.\\
{\bf Figure S1.} Results of test for combination modes from CCSM4 input data.\\
{\bf Figure S2.} Results of test for combination modes from HadISST input data.\\
{\bf Animation S1.} Evolution of SST anomalies (in K) reconstructed from NLSA modes derived from CCSM4 Antarctic SST input data. \\
{\bf Animation S2.} Evolution of SST anomalies (in K) reconstructed from NLSA modes  derived from CCSM4 Indo-Pacific SST input data. \\
{\bf Animation S3.} Evolution of SIC anomalies (dimensionless) reconstructed from NLSA modes derived from CCSM4 Antarctic SIC input data. \\
{\bf Animation S4.} Evolution of SST anomalies (in K) reconstructed from coupled NLSA modes derived from HadISST Antarctic SST and SIC input data. \\
{\bf Animation S5.} Evolution of SST anomalies (in K) reconstructed from coupled NLSA modes derived from HadISST Indo-Pacific SST input data. \\
{\bf Animation S6.} Evolution of SIC anomalies (dimensionless) reconstructed from coupled NLSA modes derived from HadISST Antarctic SST and SIC input data. 

\bibliographystyle{wileyqj}
\bibliography{ACW}

\end{document}

\maketitle

\section*{Contents}
\begin{enumerate}
\item Text S1
\item Text S2
\item Figures S1 to S2
\end{enumerate}

\section*{Additional Supporting Information (Files uploaded separately)}

\begin{enumerate}
\item Captions for Animations S1 to S6
\end{enumerate}

\section*{Introduction}

\begin{figure}
\centering
\includegraphics[width=.7\linewidth]{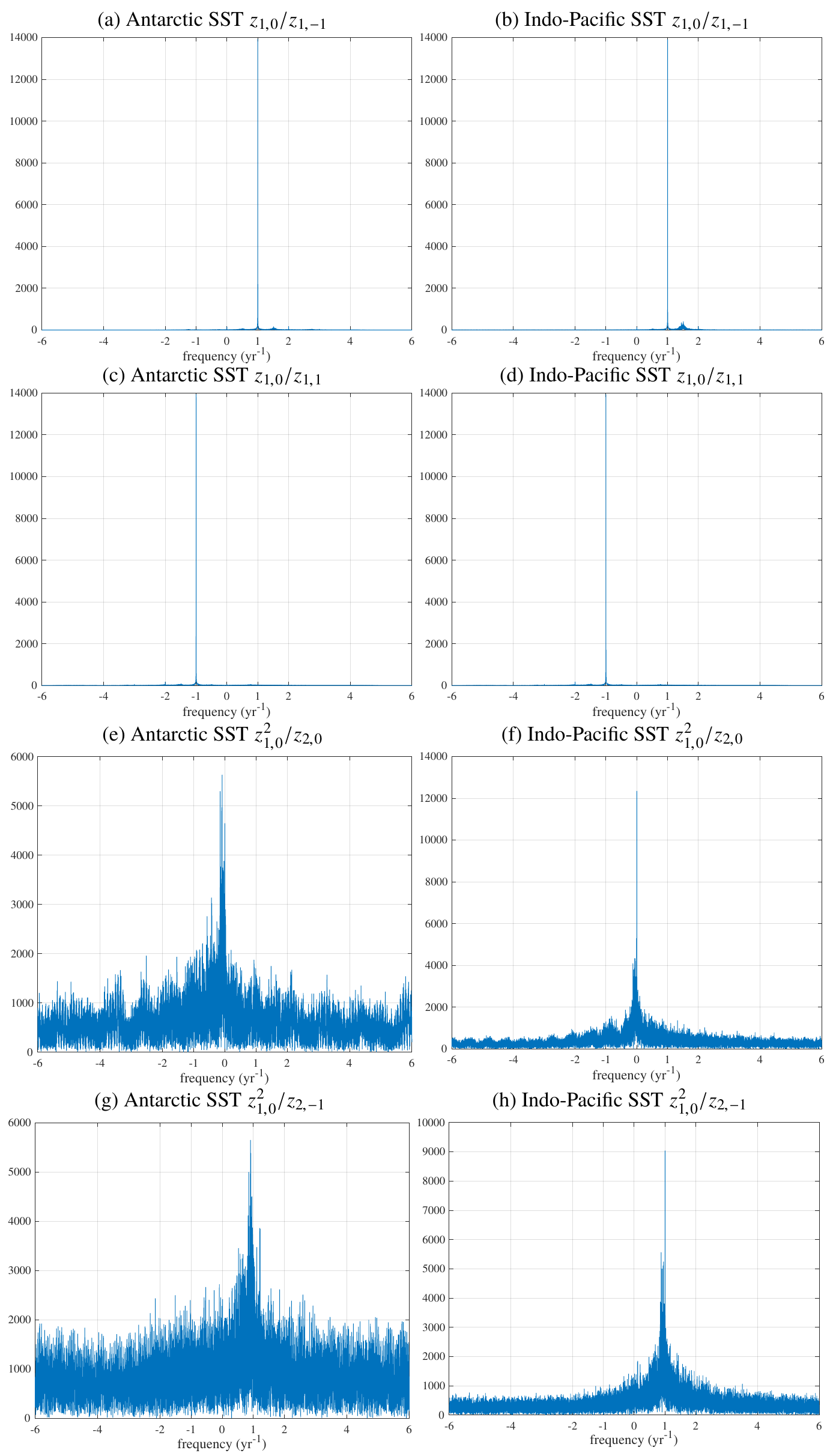}
\caption{Periodograms of representative ratios $ r_{mn} $ for wavenumber-2 ACW (left-hand column) and ENSO (right-hand column) combination modes recovered by NLSA from CCSM4 Antarctic and Indo-Pacific SST data, respectively. Notice the strong spectral peaks in (a--d) at the theoretically expected frequencies for $ r_{mn} $ for the ACW$\pm$A (a, c) and ENSO$\pm$A (b, d) combination modes. The strength of these peaks diminishes for the higher-order combination modes in (e--h), especially for the Antarctic-derived modes in (e, g), but nevertheless spectral peaks at the theoretical frequencies are still clearly visible.}
\label{figure1}
\end{figure}

\begin{figure}
  \centering
  \includegraphics[width=.8\linewidth]{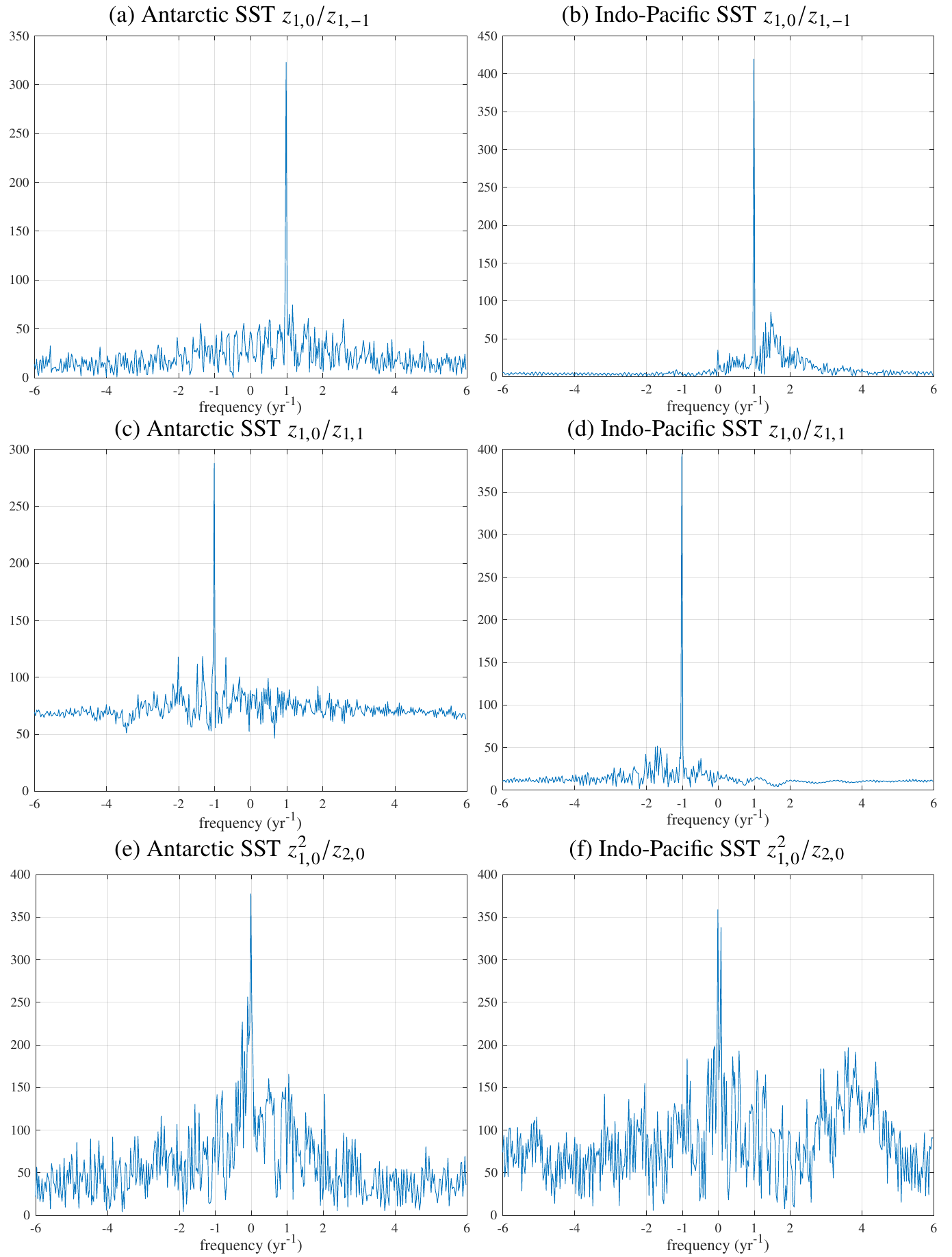}
  \caption{As in Figure~S2, but for modes derived from HadISST using joint Antarctic SST and SIC input data (left-hand column) and Indo-Pacific SST input data (right-hand column).}
  \label{figure2}
\end{figure}

This supporting information provides a description of NLSA (Text~S1), including the kernel and parameters used to obtain the results reported in the paper. We also describe our tests based on complex phase vectors to assess combination mode relationships (Text~S2). Figures~\ref{figure1} and~\ref{figure2} show the frequency relationships  between the primary ACW/ENSO modes and their associated combination modes derived from CCSM4 and HadISST data, respectively.

\section*{Text S1. Overview of NLSA algorithms}

Consider a dataset consisting of time-ordered samples $ x_1, x_2, \ldots, x_N $, where the $x_i $ are $ d$-dimensional vectors containing the observations (here, monthly averaged SST or SIC sampled at $ d $ gridpoints), taken at times $ t_i $ with a fixed timestep $ \tau = t_{i+1} - t_i$ (here, $ \tau= 1 $ mo).  As stated in Section~2, NLSA recovers temporal and spatial patterns from this dataset through the eigenfunctions of a Markov matrix, $ P $, formed by normalization of a kernel matrix, $ K $, which is in turn  computed from delay-embedded data. Specifically, selecting a positive integer parameter $ q $ (the number of delays) we consider the delay embedded vectors
\begin{displaymath}
  X_i = ( x_i, x_{i-1}, \ldots x_{i-q+1}), \quad X_i \in \mathbb{ R }^{qd}, \quad i\in \{ q, q+1,\dots, N \}.
\end{displaymath}

Using the $X_i $, the elements 
$ K_{ij} = \mathcal{ K }( X_{i}, X_{j} ) $ 
are computed, where $ \mathcal{K} $ is a kernel function providing a local measure of similarity between data points. Here, we use a so-called cone kernel \citep{Giannakis15,SlawinskaGiannakis17}, defined as
\begin{equation}
\label{eq:coneker}
 \mathcal{K}(X_i,X_j)=\exp\left(-\frac{\|X_i-X_j\|^2}{\epsilon\|\xi_i\|\|\xi_j\|}\sqrt{(1-\zeta \cos^2 \theta_{ij})(1-\zeta\cos^2\theta_{ji})}\right),
\end{equation}
where $ \epsilon $ is a positive bandwidth parameter controlling the rate of decay of the kernel, $\xi_i=(X_{i}-X_{i-1})/\tau$ are phase velocity vectors measuring the local time tendency of the data, $\cos\theta_{ij}=(X_i-X_j)\cdot\xi_i / (\|X_j-X_i\|\|\xi_i\|)$ captures the angle between the $\xi_i$ and the relative displacement vector $X_i -X_j$, and $ \zeta \in [ 0,  1 ) $ is a parameter controlling the influence of the angular terms. Notice in particular that $ \mathcal{K}(X_i,X_j) $ attains its maximum value, 1, when $ X_i = X_j $ and decays exponentially fast as the distance $ \| X_i - X_j \| $ increases. However, for $ \zeta \approx 1 $, the rate of decay is highly anisotropic, as pairs of data points whose relative displacement vector $ X_i -X_j $ is aligned with the time tendencies $ \xi_i $ and $ \xi_j $ are assigned higher similarity than pairs of points with relative displacement vector orthogonal to the time tendencies.  

The kernel matrix $ K $ is then normalized to form an $ N' \times N' $, $ N' = N - q + 1$, Markov matrix $ P $ using the procedure introduced in the diffusion maps algorithm for nonlinear dimension reduction \citep{CoifmanLafon06}. Specifically, $P$ is formed by the sequence of operations

\begin{displaymath}
  Q_i = \sum_{j=1}^{N'} K_{ij}, \quad \tilde K_{ij} = \frac{ K_{ij} }{ Q_i Q_j }, \quad D_i = \sum_{j=1}^{N'} \tilde K_{ij}, \quad P_{ij} = \frac{ \tilde K_{ij} }{ D_i },
\end{displaymath} 
and it is clear that the Markov property $ \sum_{j=1}^{N'} P_{ij} =1 $ holds. Moreover, it can be verified that $ P $ is similar to a symmetric matrix, which implies that its eigenvalues are real. The temporal patterns $ \phi_k $ are recovered by the eigenvectors of $ P $, ordered in order of decreasing corresponding eigenvalue, $ \lambda_k$, where $ \lambda_k \leq 1 $ by the Markov property of  $P $. 

\citet{BushukEtAl14} extended the NLSA algorithm described above to multivariate datasets through the use of product kernels of the form $ \hat {\mathcal{K}}( X_i, X_j ) = \mathcal{K}( X_i^{(1)}, X_j^{(1)} ) \times \mathcal{K}( X_i^{(2)}, X_j^{(2)} ) \times \cdots $, where the $ X^{(1)}_i, X^{(2)}_i, \ldots $, are lagged embedded vectors from different measured quantities. Note that the normalization by $ \| \xi_i \| \| \xi_j \| $ in Equation~(\ref{eq:coneker}) is important as it renders the argument of the exponential dimensionless (unit-free), allowing physically meaningful aggregations of the individual terms $ \mathcal{K}( X_i^{(k)}, X_j^{(k)} ) $ in a product kernel. Here, coupled NLSA is used in some of the analyses of the HadISST data, with  $ X^{(1)}_i $ and $ X^{(2)}_i $ assigned to SST and SIC data, respectively.   

In NLSA, the eigenvalues $ \lambda_k $ are not directly related to explained variance (though one can compute the variance explained by each mode in a post-processing step); rather, the quantities $ 1 - \lambda_k $ can be interpreted as a ``resolution'' of the corresponding $ \phi_k $ on the abstract manifold sampled by the data \citep{SlawinskaGiannakis17}. Every eigenvector  $ \phi_k =  (\phi_{1k}, \phi_{2k}, \ldots, \phi_{N'k} )^\top $ corresponds to a temporal pattern, $ \tilde \phi_k( t_i ) = \phi_{i+q-1,k} $, which can be thought of as a nonlinear analog of a principal component in EOF analysis and SSA. Moreover, the $ \phi_k $ can be used for spatiotemporal reconstruction as in those techniques; we refer the reader to, e.g., \cite{GhilEtAl02} and \cite{GiannakisMajda12a} for more details on this standard procedure.   

A key property of the NLSA patterns $ \tilde \phi_k $ is that they are individually temporally coherent, and collections such as   $ \{ \tilde \phi_0, \tilde \phi_1, \ldots, \tilde \phi_l \} $, where $ l \ll N' $, can capture multiple intrinsic dynamical timescales from the input data without requiring ad hoc preprocessing. This capability stems from a connection between NLSA and Koopman operators of dynamical systems due to delay embedding. Specifically, it can be shown that as the number of delays $ q $ increases, the eigenfunctions recovered from NLSA recover distinct frequencies associated with the point spectrum (eigenvalues) of the Koopman operator of the dynamical system generating the data \citep{Giannakis17,DasGiannakis17}. In practice, this means that individual NLSA eigenfunctions will tend to capture distinct intrinsic timescales of that dynamical system, and thus have high physical interpretability. Further, as shown in \citet{Giannakis15}, for $ \zeta \approx 1 $, the presence of the angular terms in Equation~(\ref{eq:coneker}) increases the capability of the technique to recover the slow (in this case, interannual) timescales in the spectrum of the dynamical system.

In our analyses, we use throughout the parameter value $ \zeta = 0.995 $. In the CCSM4 analyses, we use $ q = 144 $ (corresponding to a 12 yr embedding window) and $ \epsilon = 0.8 $. The corresponding parameter values in the HadISST analyses are $ q = 48 $ (4 yr embedding window) and $ \epsilon = 1.2 $. As in the sensitivity analysis in  \citet{SlawinskaGiannakis17}, we have verified that the family of ACW modes is qualitatively robust for values of $\epsilon$ in the interval $[0.8, 3]$ in CCSM4 data and $[0.8,2]$ in HadISST data, and our results are reseaonably stable under changes of $\zeta$ in the interval $[0.9,1)$.

\section*{Text S2. Tests for combination modes}

Consider the  complex numbers $ z_{mn} $ from Section~3.1, representing the amplitude and phase evolution of oscillatory pairs of NLSA eigenfunctions associated with the wavenumber-2 ACW combination modes. Our objective is to test for a temporal evolution of the form
\begin{displaymath}
  z_{mn}( t ) = a^m_{ACW}( t ) e^{i2\pi ( m f_{ACW} + n f_A) t },
\end{displaymath}
as suggested by combination mode theory, where $ f_{ACW} $ and $ f_A $ are the primary wavenumber-2 ACW and annual cycle frequencies, respectively, and $ a_{ACW}(t) $ a low-frequency (decadal) amplitude envelope. If this relationship holds accurately, then the ratio $ r_{mn}( t ) = z^m_{10}(t) / z_{mn}(t) $ should be a nearly pure Fourier mode at the Fourier frequency $ - n f_A $. Correspondingly, the power spectral density of $ r_{mn} $ should exhibit a strong peak at that frequency. In Figures S1 and S2, we show periodograms for various such ratios and their counterparts computed from ENSO combination modes. All periodograms were computed via standard discrete Fourier transforms.

\section*{Animation S1.} 

Evolution of SST anomalies (in K) reconstructed from NLSA modes derived from CCSM4 Antarctic SST input data. The panels in the top row show, from left to right, reconstructions based on the primary wavenumber-2 ACW modes and the  ACW$\pm$A combination modes. The panels in the bottom row show joint reconstructions based on the wavenumber-2 ACW and ACW$\pm$A modes (i.e., the sum of the reconstructions in the top row), as well as reconstructions based on the quadratic wavenumber-2 2ACW modes and the wavenumber-3 ACW modes.

\section*{Animation S2.} 
Evolution of SST anomalies (in K) reconstructed from NLSA modes  derived from CCSM4 Indo-Pacific SST input data. The panels in the top row show, from left to right, reconstructions based on  the primary ENSO modes and the  ENSO$\pm$A  combination modes. The panels in the bottom row show joint reconstructions based on the ENSO and ENSO$\pm$A modes, as well as reconstructions based on the quadratic 2ENSO and 2ENSO-A modes.

\section*{Animation S3.} 
Evolution of SIC anomalies (dimensionless) reconstructed from NLSA modes derived from CCSM4 Antarctic SIC input data. The panels show, from left to right, reconstructions based on the primary wavenumber-2 ACW modes, the ACW$\pm$A combination modes, the wavenumber-2 ACW and ACW$\pm$A modes (jointly), and the wavenumber-3 ACW modes.

\section*{Animation S4.} 
Evolution of SST anomalies (in K) reconstructed from coupled NLSA modes derived from HadISST Antarctic SST and SIC input data. The panels in the top row show, from left to right, reconstructions based on the primary wavenumber-2 ACW modes and the ACW$\pm$A combination modes. The panels in the bottom row show joint reconstructions based on the wavenumber-2 ACW and ACW$\pm$A modes, as well as reconstructions based on a pair of modes that display a mixture of wavenumber-2 2ACW and wavenumber-3 ACW characteristics, and a pair of ACW$-$2A semiannual combination modes. 

\section*{Animation S5.}
Evolution of SST anomalies (in K) reconstructed from coupled NLSA modes derived from HadISST Indo-Pacific SST input data. The panels in the top row show, from left to right, reconstructions based on the primary ENSO modes and the ENSO$\pm$A combination modes. The panels in the bottom row show joint reconstructions based on the ENSO and ENSO$\pm$A modes, as well as reconstructions based on the quadratic 2ENSO modes, and a pair of ENSO$-$2A semiannual combination modes. .
 
\section*{Animation S6.} 
Evolution of SIC anomalies (dimensionless) reconstructed from coupled NLSA modes derived from HadISST Antarctic SST and SIC input data. The panels show, from left to right, reconstructions based on the primary wavenumber-2 ACW modes, the ACW$\pm$A combination modes, the wavenumber-2 ACW and ACW$\pm$A modes (jointly), and a pair of modes displaying a mixture of wavenumber-2 2ACW and wavenumber-3 ACW features.

\bibliographystyle{wileyqj}
\bibliography{ACW}

